\documentclass[aps,showpacs,superscriptaddress,twocolumn]{revtex4-1} 
\usepackage{epsfig} 
\usepackage{amsmath,amssymb,amsthm} 
\usepackage{graphicx}
\usepackage{psfrag}
\usepackage{bm}

\def\s2{\sigma^2}  
 
\begin{document}

\title{Wave packet revivals in a  graphene quantum dot in  a perpendicular 
  magnetic field} 
 
\author{J. J. Torres }
\affiliation{Instituto Carlos I de F{\'\i}sica Te\'orica y
Computacional, Universidad de Granada, Fuentenueva s/n, 18071 Granada,
Spain}
\affiliation{Departamento de Electromagnetismo y F{\'\i}sica de la 
Materia, 
Universidad de Granada, Fuentenueva s/n, 18071 Granada, 
Spain}  
\author{ E. Romera } 
\affiliation{Instituto Carlos I de F{\'\i}sica Te\'orica y 
Computacional, Universidad de Granada, Fuentenueva s/n, 18071 Granada, 
Spain} 
\affiliation{Departamento de F\'isica At\'omica, Molecular y Nuclear, 
Universidad de Granada, Fuentenueva s/n, 18071 Granada, Spain 
}

\date{\today}  
  
\begin{abstract}  
We study the time-evolution of localized wavepackets in graphene quantum dots 
in a perpendicular magnetic field, focusing on the quasiclassical and 
revival periodicities, for different values of the magnetic field intensities 
in a theoretical framework. We have considered contributions of the two 
inequivalent points in the Brillouin zone. The revival time has been found 
as an observable that shows the break  valley degeneracy.     
 \end{abstract}  
\maketitle  
  
\section{Introduction}  
Graphene, a single sheet of graphite, has attracted growing interest due to 
its remarkable and starling properties and its potential applications in 
nanoelectronics \cite{castro,Geim,Novoselov,mech,resistivity}. Many of these 
properties are attributed to the existence of a peculiar band structure 
constituted by quasifree electrons being described by a massless 
Dirac equation \cite{massless}. In fact, perfect tunneling of electron wave packets through potential barriers (the Klein paradox) \cite{kp} or the unconventional 
quantum Hall effect, have been experimentally confirmed in graphene 
\cite{qhe}. It has been also argued that graphene in a magnetic field is a 
promising system for the experimental observation of the highly  oscillatory 
motion of the electric current known as zitterbewegung 
\cite{rusin1,rusin2,rusinnuevo,editsug}. In addition, revivals of electric currents in 
graphene in the presence of a magnetic field \cite{romera} have been 
theoretically predicted  and  an exhaustive study of quantum wave-packet 
revivals, fractional revivals, classical periodicity in graphene was  reported \cite{romera,kramer}.  
 
 In recent times, graphene quantum dots have been widely studied, both 
theoretically and experimentally. It has been reported that there are different 
theoretical approaches which allow to electrostatically define a graphene 
quantum dot, or to create it by cutting flakes of graphene 
\cite{21,22,23,24,25,26,schnez,32}. Also, there has been an analysis of 
the level  
statistics of the graphene quantum dots observed in 
experiments (for quantum dots  smaller than 100 nm) which has concluded that 
it is well described by the theory of 
chaotic Dirac neutrino billiards \cite{ponomarenko}. On the other hand, new experiments have 
been designed for the creation of graphene quantum dots \cite{34,35,36} by etching or scratching graphene islands including the study of 
energy levels in magnetic fields \cite{36,39}.  
 Additionally, 
graphene quantum dots have been point out to be very attractive as spin qubits in quantum information processing \cite{recher1}.

 The temporal evolution of wavepackets, relativistic and nonrelativistic, 
 displays interesting revival phenomena due to quantum interference. Several 
 types of periodicity may emerge depending on the energy eigenvalue spectrum, 
 in this regard revivals and fractional revivals have raised great interest 
 during recent years \cite{1}. Propagating wave packets initially evolve 
 quasiclassically and oscillate with a period $T_{cl}$, and later the wavepackets 
 spread and collapse. For longer times,  the wavepackets evolve in the so-called 
 revival time $T_{rev}$ to a state that closely reproduces its initial 
 waveform. Fractional revivals appear as a temporal self-splitting of the 
 initial wavepacket into a collection of a number of scaled copies   at 
 times  $t=p T_{rev}/q$, with $p$ and $q$ mutually 
 prime. Assuming that the initial wavepacket is a superposition of the 
 eigenstates $u_n$  sharply peaked around some $n_0$, revival times can  
 be obtained from the Taylor series of the energy spectrum $E_n$ around 
 $E_{n_0}$ as $T_{rev}=4\pi \hbar/|E_{n_o}^{\prime\prime}|$. Revivals  have been 
 investigated theoretically and experimentally in nonlinear quantum systems 
 \cite{1,1b,1c,rob,yeazell,exp}. Recently, revivals of electric current in monolayer graphene in the presence of 
 an external magnetic field have been also described \cite{romera,kramer}. 
 
This article is organized as follows. In section II the phenomenon of 
wavepacket revivals in a graphene quantum dot in a magnetic field using a 
model Hamiltonian is studied.   
 We discuss on the relevance of quantum revivals for the 
characterization of the breaking of the valley degeneracy in graphene quantum 
dots.

\section{Wave packet revivals for graphene quantum dots in a magnetic field}  
  
To study wave packet revivals in a graphene quantum dot in a   
perpendicular magnetic field  
 ${\bf B}$ we consider the  
model Hamiltonian for electrons in the valley-isotropic form which is given by \cite{schnez}  
\begin{equation}  
H_{\tau}=v_F({\bf p} + e{\bf A})\cdot {\boldsymbol \sigma} + \tau 
V(r)\sigma_z. 
\label{ham}  
\end{equation}   
We use the symmetric gauge for the vector potential, ${\bf  
  A}=B/2(-y,x,0)=B/2(-r\sin\phi,r\cos\phi,0)$, where $\phi$ the polar  
angle, $v_F=10^6 m/s$  the Fermi velocity, and   $\tau=\pm 1$ differentiates the two  
valleys $K_1$ and $K_2$. ${\boldsymbol \sigma}$ are Pauli's spin matrices in the  
basis of the two sublattices of A and B atoms. We assume the confinement  
potential in \cite{schnez}, which is  
a mass-related potential energy $V(r)$ coupled to the Hamiltonian via the  
$\sigma_z$ Pauli matrix, where $V(r)=0$ for $r<R$ and $V(r)=\infty$ for $r>R$,  
that is, tends to infinity in the edge of the dot.  
 
\begin{figure}[t!] 
\begin{center} 
\includegraphics[width=8.5cm]{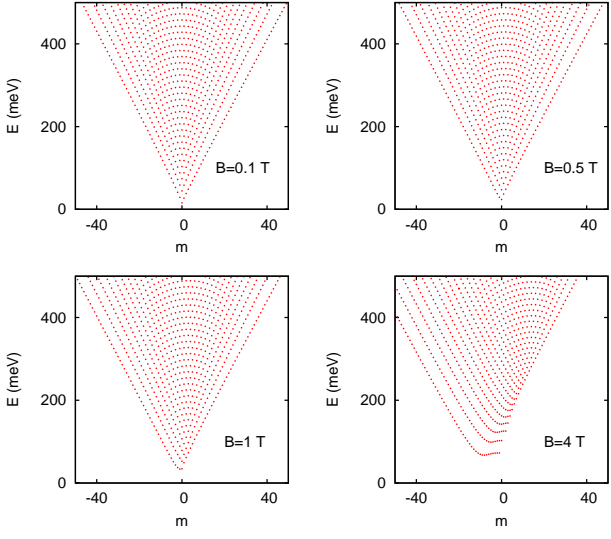} 
 
\caption{Positive energy spectrum for a graphene Quantum Dot, with $R=70~nm,$ versus 
  the quantum number $m$ and different values of 
  the magnetic field $B$ ($K_1$ point).} 
\label{fig1} 
\end{center} 
\end{figure} 
 
\begin{table} 
\begin{tabular}{|l|ll|ll|} 
\hline 
\hline 
$B (T)$ & $T_{cl}$ (ps) & $T_{rev}$ (ns) & $T^{\prime}_{cl}$ (ps) & 
$T_{rev}^{\prime}$ (ns)\\ 
\hline 
0.1  &  0.14 &   115.8 & 0.14 & 120.7 \\ 
0.5 & 0.14 &6.8 & 0.14 & 7.6 \\ 
1  & 0.14 & 2.3 & 0.14 &2.6\\ 
4& 0.14& 0.2 & 0.14 & 0.3 \\ 
\hline 
\hline

\end{tabular} 
\caption{Classical period $T_{cl}$ and 
  revival time $T_{rev}$ in the two valleys ($K_1$ and $K_2$) in the band 
  structure of  graphene quantum dot for different values of the 
magnetic field $B$ and for an initial wave packet with $n_0=20$, $m=0$, $\sigma=1.1$.} 
\label{tab1} 
\end{table}  
Upon introducing the magnetic length $l_B=\sqrt{\hbar/(eB)}$ and using the  
fact that $H_{\tau}$ commutes with the total angular momentum operator  
$J_z=L_z+\frac{\hbar}{2}\sigma_z$, $[H_{\tau},J_z]=0$, the solution of the  
Dirac equation $H_{\tau}\psi(r,\phi)=E\psi(r,\phi)$ (where  
$\psi(r,\phi)=[\psi_1(r,\phi),\psi_2(r,\phi)]$ is a two-component spinor) is  
given by \cite{schnez}:  
\begin{equation}  
\begin{array}{ll} 
\displaystyle\psi_1(r,\phi)&\displaystyle=c e^{im\phi} r^m e^{-r^2/4l_B^2}  
 L\left(\frac{k^2l_B^2}{2}-(m+1),m,r^2/2l_B^2\right) \\ 
\\  
\displaystyle \psi_2(r,\phi)&\displaystyle=c e^{i(m+1)\phi} r^m e^{-r^2/4l_B^2}\\ 
\\   
\displaystyle &\displaystyle\times  \frac{r/l_B}{kl_B}L\left(  
  \frac{k^2l_B^2}{2}-(m+1),m+1,r^2/2l_B^2\right)  
\end{array} 
\label{chara} 
\end{equation} 
where $L(a,b,z)$ is the generalized Laguerre function and $c$ is a normalization constant.  
The characteristic equation for the allowed eigenenergies $E$ of the  
Quantum Dot using the boundary condition  for a circular confinement  
\cite{berry,schnez} $\psi_2/\psi_1=\tau \exp{[i\phi]}$ has been obtained in   
\cite{schnez} and it can be written as 
\begin{equation}  
\begin{array}{l} 
\displaystyle \left(1-\tau\frac{kl_B}{R/l_B}\right)  
L\left(\frac{k^2l_B^2}{2}-(m+1),m,R^2/2l_B^2\right)\\  
 \\ 
 \displaystyle + L\left(\frac{k^2l_B^2}{2}-(m+2),m+1,R^2/2l_B^2\right)=0.  
\end{array} 
\label{cond1} 
\end{equation} 
\begin{figure}[ht!] 
\begin{center} 
\includegraphics[width=9cm]{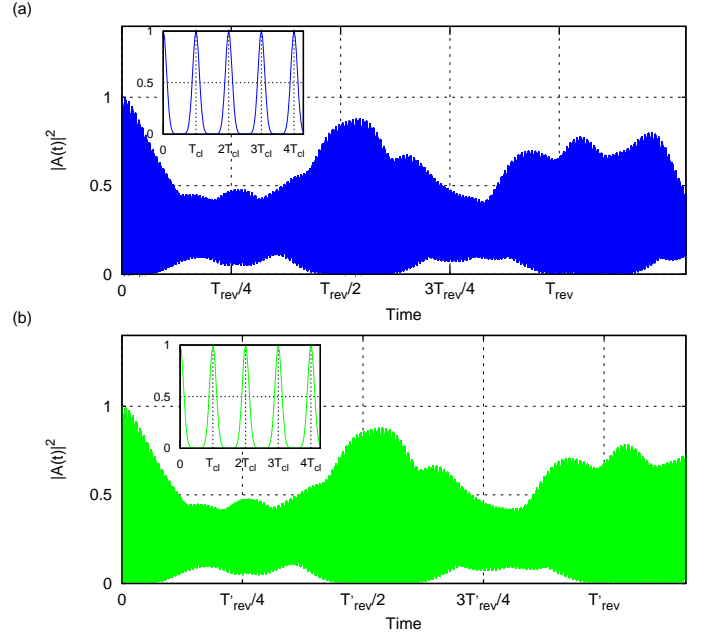} 
\caption{Time dependence of $|A(t)|^2$ for an initial spherically symmetric  Gaussian wavepacket 
  with $n_0=20$, $\sigma=1.1$,  $B=0.2$ T and $R=70$ $nm$ for (a) the valley 
  $K_1$ ($T_{rev}=28$ ns) and (b) the valley $K_2$ ($T^{\prime}_{rev}=31$ ns).} 
 
\label{fig2} 
\end{center} 
\end{figure} 
 
 
The spectrum of electrons confined in a circular graphene quantum dot in a  
perpendicular magnetic field is given by $E(n,m,\tau)$ verifying (\ref{chara}) 
and (\ref{cond1}) 
and the energy eigenfunctions are $\psi_{(n,m)}(r,\phi)$. Schnez et  
al. \cite{schnez} have analyzed the limits for $B=0$ and  
$R/l_B\rightarrow\infty$ cases. For $B=0$ the eigenenergies are given by the  
characteristic equation $\tau J_m(kR)=J_{m+1}(kR)$. On the other hand,  
they showed  that the energy levels converge to the  
 landau levels $E_m=\hbar v_F k_m=\pm v_F\sqrt{2e\hbar B (m+1)}$ for 
 $R/l_B\rightarrow\infty$ .

The initial wave packets will be constructed  as a superposition of eigenstates 
assuming a Gaussian population of energy levels  
\begin{equation} 
\Psi(r,\phi)=\sum_{n,m} c_{n,m} \psi_{(n,m)}(r,\phi) 
\end{equation}  
 that is, we choose $c_{n,m}$  Gaussianly  distributed and 
centered around some ${n_0,m_0}$ with width $\sigma$ (See for instance 
Refs. \cite{rob,banerji06,bluhm96}). We will study  wave packet 
dynamic of this initial wave packet for different values of the intensity 
of the magnetic field $B$.    The particular assumption we have made to 
 construct the initial wavepacket has an influence in the  results (see 
 i. e. \cite{rob}). Moreover, 
  the construction of a realistic initial wavepacket (i.e. by laser 
 excitation) would require further considerations.

We start  studying the special  case of a radially 
symmetric wave packet, that is, a superposition of purely $m=0$ states. For an initial wave packet sharply peaked around a central $n_0$, the 
different time scales can then be identified from the coefficients of the Taylor 
expansion of the energy spectrum $E_n$ around $E_{n_0}$: 
\begin{equation} 
E_{n}=E_{n_0}+ \frac{2\pi\hbar}{T_{cl}} (n-n_0) + \frac{4\pi\hbar}{T_{rev}} (n-n_0)^{2}+... 
\end{equation}  
In Table I (middle column) the times scales 
$T_{cl}$ and $T_{rev}$ have been  numerically fitted from energies of Fig. 1.  
 The quantum number 
dependence of the energy eigenvalues is  necessary to determine  the 
classical period and the revival time  in a system.  In 
Fig. \ref{fig1} we plot  the $n$ and $m$ dependence of the energies $E_{n,m}$ for $B=0.1$, $0.5$, $1$ and $4$ 
T and  
for $R=70$ nm in the valley $K_1$.  
 
To analyze the evolution of the wave packet  we shall use  the autocorrelation function 
\begin{equation} 
 A(t)=\int r\,dr\,d\phi \; \psi^*(r,\phi,0) \psi(r,\phi,t). 
\end{equation} 
An alternative approach based on information entropies has been recently 
proposed \cite{romera1}. 
The occurrence of revivals correspond to the approximate return of $|A(t)|^2$ 
to its initial value of unity. 
 The time evolution of the autocorrelation function for an initial wave packet 
 with $n_0=20$, $m=0$ 
$\sigma=1.1$, $R=70$ $nm$ and $B=0.2$ $T$ is showed  in Fig. \ref{fig2} (top panel).  We can see that at early times the 
Gaussian wave packet evolves quasiclassically in the inset of the figures. For longer time scales  the 
wave packet initially spreads and delocalizes. A revival of the wavepacket can 
be observed for $T_{rev}/2$ (mirror revival \cite{rob}). 
 
\begin{figure}[ht!] 
\begin{center} 
\includegraphics[width=8.5cm]{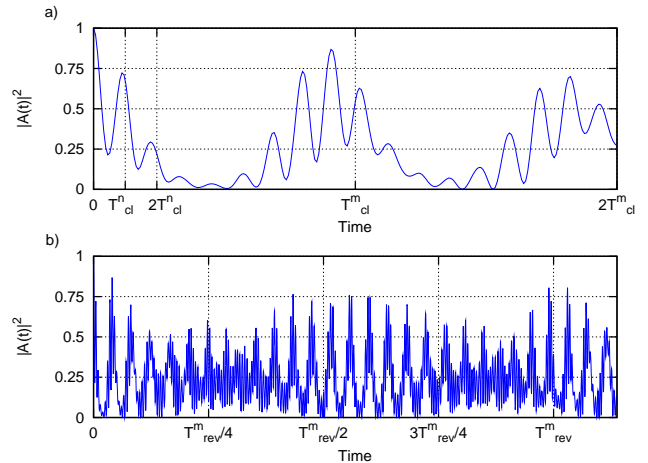} 
\caption{Time dependence of $|A(t)|^2$ for initial wavepackets with $B=4$ T, 
  $R=70 ~nm$, $n_0=1$, $m_0=-15$ and $\sigma_n=\sigma_m=0.7$ in the valley $K_1$.(a) 
  First classical periods of motion with $T_{cl}^{(n)}=0.08 ~ps$, 
  $T_{cls}^{(m)}= 0.65 ~ps$ (b) 
  Long time dependence with $T_{rev}^{(n)}=2.09 ~ps$ and $T_{rev}^{(m)}=14.50 ~ps$.} 
\label{fig4} 
\end{center} 
\end{figure}

Let us turn to examine quantum revivals in the two inequivalent points in 
the Brillouin zone $K_1$ and  $K_2$ which corresponds with Hamiltonian 
(\ref{ham}) with $\tau=1$ and  
$\tau=-1$, respectively. We have calculated the classical and revival times for different 
values of the magnetic field $B$ which are summarized in  table 
\ref{tab1}. The quasiclassical evolution is analogous in both valleys as we can 
see  in the  insets of Fig. \ref{fig2}. We 
  have found both theoretically and by simple inspection of the behavior of the autocorrelation 
  function in Fig. \ref{fig2}, that $T_{rev}$ differs in both zones clearly 
  for a  
  magnetic field $B=0.2 ~T$.  In fact,  it is known that the valley degeneracy is 
broken by the perpendicular magnetic field \cite{recher1,26}. 
 So, we have found the 
revival time as  an observable which shows explicitly the breaking of valley 
degeneracy in a graphene quantum dots.  We would like to point 
out that the 
breaking of the valley degeneracy is a prerequisite for spin-based quantum 
computing in graphene quantum dots . 
 As the magnetic field is growing the revivals times are 
  equalizing  as we expect for Landau levels regime  \cite{romera}. 
\begin{figure}[ht!] 
\begin{center} 
\includegraphics[width=8.5cm]{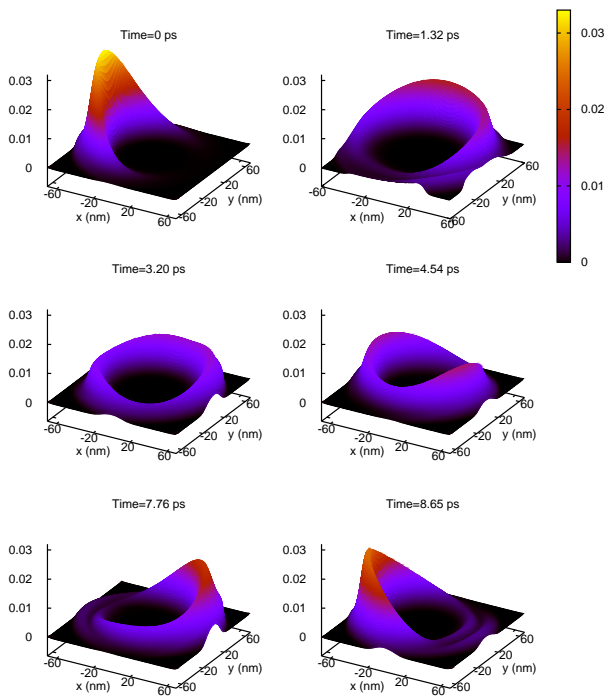} 
\caption{Snapshots of the probability density $|\psi(x,y)|^2$ for an initial  gaussian wavepacket with $n_0=1$, $m_0=-15$, 
  $\sigma_n=\sigma_m=0.7$, $B=4 ~T$ and $R=70 ~nm$. It is  noticeable the 
  existence of fractional revivals after the delocalization of the initial 
  wave packet (middle panels) and the revival at $t=8.65 ~ps$.} 
\label{fig5} 
\end{center} 
\end{figure}

Finally, we  consider an initial superposition state with coefficients taken from a Gaussian distribution 
centered around $m=-15$ and $n=1$ for $B=4T$. In this case the energy 
levels are characterized by  
two quantum numbers $E(n,m)$. Systems with two quantum numbers offer richer 
possibilities for wave packet revivals (for a theoretical analysis see 
\cite{rob}). The classical periods yield straightforwardly 
$T_{cl}^{n}=0.078~ps$ and $T_{cl}^{m}=0.65~ps$ (from \cite{rob} 
$T_{cl}^{(n)}=2\pi\hbar/\left|\frac{\partial E}{\partial 
      n}\right|_{\tiny n_0,m_0}$ and 
  $T_{cl}^{(m)}=2\pi\hbar/\left|\frac{\partial E}{\partial 
        m}\right|_{n_0,m_0}$, respectively). Now, the long-time revival 
    structure depends on three possible times, given by \cite{rob} 
$ 
T_{rev}^{(n)}=4\pi\hbar/\left|\frac{\partial^2 E(n,m)}{\partial 
      n^2}\right|_{n_0,m_0}$,  
$T_{rev}^{(m)}=4\pi\hbar/\left|\frac{\partial^2 E(n,m)}{\partial 
      m^2}\right|_{n_0,m_0}$ and   
$T_{rev}^{(n,m)}= 2\pi\hbar/\left|\frac{\partial^2 E(n,m)}{\partial 
      n\partial m}\right|_{n_0,m_0}$, and the revival structure is constructed 
  on the interplay between these times.  
This is shown in Fig. \ref{fig4} for 
  $B=4T$. In the top panel it is apparent that at short times the 
  quasiclassical oscillatory behavior with period $T_{cl}^{(n)}=0.078~ps$. At 
  medium times the quasiclassical behavior corresponding to 
  $T_{cl}^{(m)}=0.65~ps$ is superimposed on the first oscillations. 
 In the bottom panel revivals can be seen at 
  $t\approx 8.5 ps$ as an interplay between the revival times, as we would expect. Snapshots of the numerical simulation of the probability density 
  are given in Fig. \ref{fig5} at several times. It is apparent that the regeneration 
  of the density function is given at time $t=8.65~ps$. 
 
 We want to remark that 
the typical scattering times should be taken into account for a more realistic 
description of this problem. 
Moreover, it would be interesting to understand the influence of shape 
deformations, impurities, or interactions on revivals, so a more detailed 
study of this topic could be made in a analogous way that has been done before (see 
i.e. \cite{romera,kramer}) in 
graphene in a perpendicular magnetic field.

\section{Conclusions}  
  
In this work  we have presented  time-dependent effects in the propagation of 
wavepackets on  graphene quantum dots for different values of the magnetic 
field. When the wavepacket is sufficiently localized around some central quantum 
 numbers ($n_0$, $m_0$), it evolves quasiclassically. At later times, it spreads 
 leading to the collapse of the classical oscillations, but at times  that are 
 multiple or rational fractions of $T_{rev}$, the wavepacket regains its 
 initial form. 
 We have also shown revival time as an observable  
 sign of breaking the valley degeneracy, that it is known as a prerequisite 
 for spin-based quantum computing. 
 It can be  underline that we have found revivals of the electron 
 wavepacket in a Dirac  system and  we suggest that this   realistic system could be 
 used  to  simulate 
   relativistic 
 quantum revivals \cite{romera,strange}.   
\section{Acknowledgments}  
 We want to thank  J. A. L\'opez-Villanueva and F. de los Santos for fruitful discussions. 
This work was supported by projects FQM-165/0207, FIS2008-01143, FQM-01505 and FIS2009-08451.

\end{document}